\documentclass[aps,prl,preprint,groupedaddress,superscriptaddress, 
nobibnotes,nofootinbib,showkeys]{revtex4-1}
\usepackage{amsmath,amssymb,graphicx}
\usepackage{float}
\usepackage{slashed}
\usepackage[title]{appendix}
\begin{document}

\preprint{\parbox[b]{1in}{ \hbox{\tt PNUTP-22/A05} }}

\title{ Unitarity and Dilaton effective theory}

\author{Deog Ki Hong}
 \email[E-mail:\,]{dkhong@pusan.ac.kr}
\affiliation{Department of
Physics,   Pusan National University,
             Busan 46241, Korea}          

  \author{Gyurin Kim}
\affiliation{Department of
Physics,   Pusan National University,
             Busan 46241, Korea} 
             
  \author{Jun Beom Park}
\affiliation{Department of
Physics,   Pusan National University,
             Busan 46241, Korea}


\date{\today}

\begin{abstract}
From the low-energy effective theory of dilatons, consistent with the scale anomaly, we calculate the $2\to2$ scattering amplitudes of dilatons. We find that the one-loop amplitude violates the unitarity bound as the scattering energy  approaches the cutoff of the effective theory, $\sqrt{s}=\sqrt{4\pi} f_D$. We then show that the inclusion of the next-to-lightest state, namely the spin-2 state, of mass around the cutoff improves the unitarity. The unitarity argument suggests that the mass ratio of the dilaton and the spin-2 state is proportional to the square of the Miransky-BKT scaling of the near conformal dynamics. 
	
\end{abstract}


\maketitle

\subsection{Introduction}
In the physics of elementary particles we observe a wide separation of scales so that at each scale a finite set of operators effectively describes their physical processes, but often the microscopic origin of the scale separation remains unanswered.  A well-known example is the electroweak theory that describes the electroweak process extremely well with no sign of its deviation at colliders. The origin of the electroweak scale or the vacuum expectation value of the Higgs field, $v_{\rm ew}=246~{\rm GeV}$, 
is however yet to be uncovered. 

One of the attractive ideas to explain the vast hierarchy of scales in quantum field theory is to generate it dynamically. A prominent example is quantum chromodynamics (QCD). 
Classically, QCD is scale-invariant, but 
the quantum effects create dynamically the scale, $\Lambda_{\rm QCD}$, 
by the so-called dimensional transmutation of the running coupling, which is determined experimentally to be about $220~{\rm MeV}$.  Furthermore, 
in the infrared (IR) the gauge coupling becomes so strong that QCD undergoes a confinement phase transition, generating the vacuum energy, ${\cal E}_{\rm vac}\sim -\Lambda_{\rm QCD}^4$~\cite{Campostrini:1989uj}.
The salient feature of the spontaneous generation of vacuum energy in the (quasi) scale-invariant theory is the appearance of a light dilaton in the low-energy spectrum~\cite{Ellis:1984jv,Yamawaki:1985zg,Bando:1986bg}, namely the Nambu-Goldstone (NG) boson associated with the spontaneous breaking of scale symmetry. 
Recently a model of near conformal dynamics that generates a large separation of scales is proposed to explain naturally the light Higgs and at the same time the dark matter  with very light dilaton~\cite{Hong:2017smd}. 

If the near conformal dynamics is responsible for light Higgs, there will be other light states in addition to the dilaton, which  might exhibit interesting signatures at colliders~\cite{Davoudiasl:2009cd,Agashe:2020wph}. Recent lattice study shows that the ground-state glueball behaves like a dilaton, exhibiting a universal scaling law in the confining gauge theories~\cite{Hong:2017suj}. It further shows that the next light state in pure Yang-Mills (YM) theories is the spin-2 glueball, $2^{++}$,  whose mass may be related to the spin-0 ground-state glueball or the dilaton in a universal way for all gauge groups~\cite{Bennett:2020hqd}. 
In this paper, we calculate the one loop scattering amplitudes of dilatons and show how the violation of the perturbative unitarity of the scattering amplitude is improved with additional spin-2 massive particles till the energy approaches the mass of the spin-2 states. We find that the unitarity argument for the new heavier state requires the mass of the spin-2 state should be at the order of the dilaton decay constant, which is consistent with the universal behavior of mass predicted in~\cite{Bennett:2020hqd}. The mass ratio between two lowest-lying states hence measures the degree of the conformality of their microscopic theory.

\subsection{Dilaton scattering amplitudes}
When the scale symmetry is spontaneously broken, as in the confining phase of YM theories, the dilatation current creates the dilaton, $\sigma$, out of vacuum by the Goldstone theorem: 
\begin{equation}
\left<0\right|{\cal D}_{\mu}\left|\sigma(p)\right>=-if_Dp_{\mu}e^{-ip\cdot x}\,,
\end{equation}
where $f_D$ is the dilaton decay constant and ${\cal D}_{\mu}$ is the dilatation current. Because of the scale anomaly, the dilatation current is not conserved,
\begin{equation}
\partial^{\mu}\left<{\cal D}_{\mu}\right>=\left<T_{\mu}^{\mu}\right>=4\,{\cal E}_{\rm vac}\,,
\label{vac}
\end{equation}
leading to mass or potential energy to the dilaton. The low energy effective Lagrangian for the dilaton, that is consistent with the (anomalous) scale symmetry, can be written as, keeping only the lowest number of derivatives,  
\begin{equation}
{\cal L}_{\rm eff}=\frac12\partial_{\mu}\chi\partial^{\mu}\chi-V_A(\chi)\,,
\label{eff}
\end{equation}
where $\chi=f_De^{\sigma/f_D}$ describes the small fluctuations of the order parameter of the scale symmetry, namely the trace of the energy-momentum tensor, around the vacuum defined in~Eq.\,(\ref{vac}),
\begin{equation}
T_{\mu}^{\mu}\approx 4\,{\cal E}_{\rm vac}\left(\frac{\chi}{f_D}\right)^4\,.
\end{equation}
Being a NG boson, the dilaton transforms nonlinearly under the scale transformation, $x\to x^{\prime}=e^{\alpha}x$, 
\begin{equation}
\sigma\to \sigma+\alpha\,f_D\,,
\end{equation}
the dilaton potential, not invariant under the scale transformation, is generated by the scale anomaly and also by other possible explicit symmetry-breaking terms in the original theory among which we ignore the latter for the current discussions.

Matching the scale anomaly, Eq.\,(\ref{vac}), one finds the dilaton potential to be~\cite{Migdal:1982jp}
\begin{equation}
V_A(\chi)=\left|{\cal E}_{\rm vac}\right|\left(\frac{\chi}{f_D}\right)^4\left[4 \ln\left(\frac{\chi}{f_D}\right)-1\right]\,.
\end{equation}
Expanding the effective Lagrangian,~Eq.\,(\ref{eff}), in powers of $\sigma/f_D$, one gets therefore 
\begin{equation}
{\cal L}_{\rm eff}=\frac12\partial_{\mu}\sigma\partial^{\mu}\sigma-\frac12m_D^2\sigma^2+\frac{\sigma}{f_D}\left(\partial_{\mu}\sigma\right)^2-\frac{4m_D^2}{3f_D}\sigma^3+\frac{\sigma^2}{f_D^2}\left(\partial_{\mu}\sigma\right)^2-2\frac{m_D^2}{f_D^2}\sigma^4+\cdots\,,
\end{equation}
where the ellipsis denotes the higher order terms and the dilaton mass $m_D^2=16\left|{\cal E}_{\rm vac}\right|/f_D^2$, given by the partially conserved dilatation current (PCDC) relation~\cite{Choi:2012kx}. 

By the construction the effective Lagrangian works well at low energy, consistently with the current algebra of the microscopic theory~\cite{Coleman:1969sm,Callan:1969sn}.  As the energy increases, however, it will approach a cutoff above which the effective theory is no longer valid.  Beyond the cutoff it exhibits  unphysical behaviors  such as the violation of the unitarity in the scattering amplitudes, signaling the existence of a new state. 
To see this, let's consider the $2\to2$ dilaton scattering amplitude
\begin{equation}
\sigma(p_1)+\sigma(p_2)\mapsto \sigma(p_3)+\sigma(p_4)\,.
\end{equation}
\begin{figure}[t]
\centering
\begin{minipage}[c]{\linewidth}
\centering
  \includegraphics[scale=0.33]{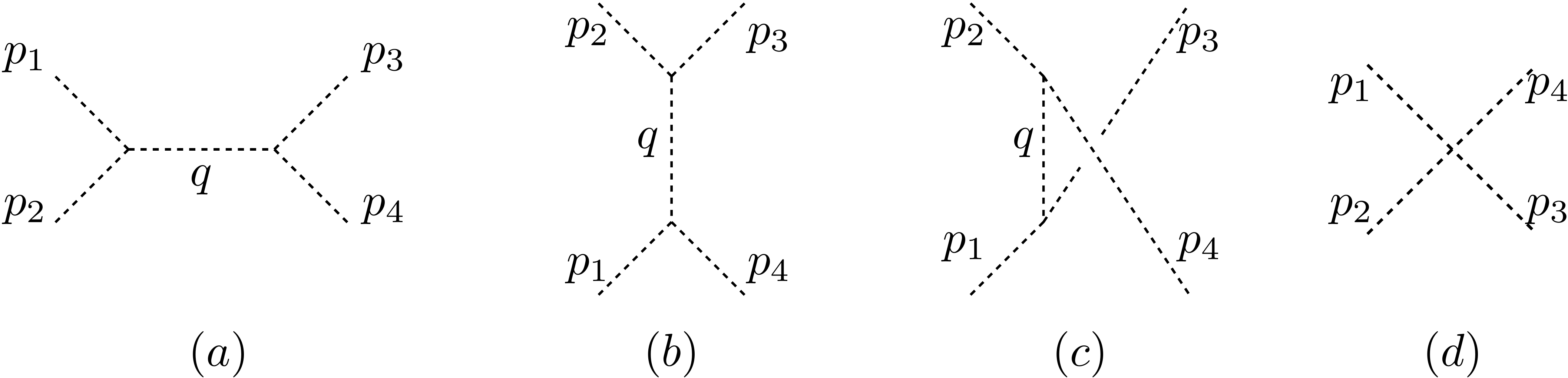}
\end{minipage}
\caption{Tree level diagrams for $2\to2$ scattering amplitude of dilatons, denoted as broken lines: (a) $s$-channel, (b) $t$-channel, (c) $u$-channel and (d) the contact term. }
\label{tree}
\end{figure}
At the tree level, shown in Fig.~\ref{tree}, the amplitude becomes
\begin{equation}
{\cal A}_{\sigma}^{\rm tree}=-\frac{1}{f_D^2}\left[\frac{(s-6m_D^2)^2}{s-m_D^2}+\frac{(t-6m_D^2)^2}{t-m_D^2}+\frac{(u-6m_D^2)^2}{u-m_D^2}
\right]\,-\frac{48 m_D^2}{f_D^2},
\end{equation}
where the Mandelstam variables $s=(p_1+p_2)^2$, $t=(p_3-p_1)^2$, and $u=(p_4-p_1)^2$\,. For high energy, $s,t,u\gg m_D^2$, the tree amplitude is found to be well behaved, since $s+t+u=4m_D^2$;
\begin{equation}
{\cal A}_{\sigma}^{\rm tree}\approx-\frac{1}{f_D^2}\left(s+t+u+15m_D^2\right)=-\frac{19m_D^2}{f_D^2}\,.
\end{equation}
\begin{figure}[H]
\centering
\begin{minipage}[c]{\linewidth}
\centering
  \includegraphics[scale=0.33]{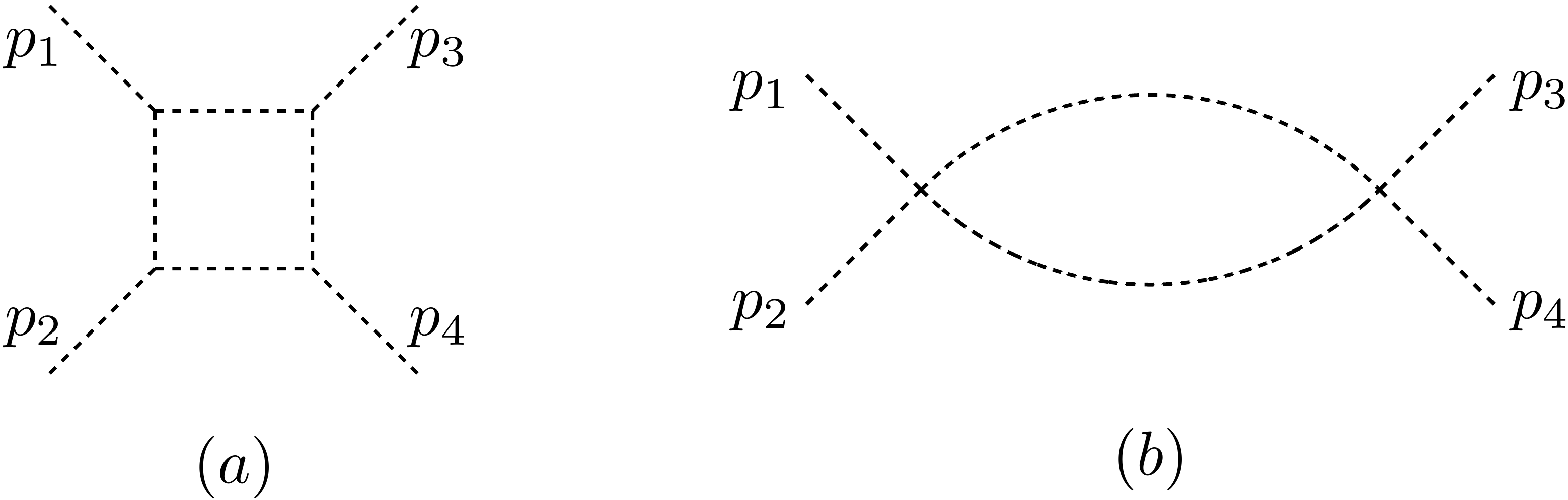}
\end{minipage}
\caption{$s$-channel  one-loop amplitude; (a) box diagram, (b) fish diagram. $t$-channel and $u$-channel amplitudes are obtained by swapping $p_3\leftrightarrow p_2$ and $p_3\leftrightarrow p_4$ for each diagram. }
\label{box}
\end{figure}
Because of the symmetry under the exchange of the external dilatons, the leading non-trivial amplitude comes from  the one-loop, which is also consistent with the scale anomaly argument in~\cite{Komargodski:2011vj}. The one-loop box diagram from Fig.~\ref{box}\,(a), adding together with its counter-parts in $t$ and $u$ channels, grows quadratically at high energy, which we have found in the minimal subtraction scheme as
\begin{equation}
{\cal A}_{\sigma}^{\rm box}=\frac{1}{32\pi^2f_D^4}\left(s^2+t^2+u^2\right)\left[2+\ln\left(\frac{4\pi^2\mu^2}{m_D^2}\right)\right]+{\cal O}\left(\frac{m_D^4}{f_D^4}\right)\,,
\label{box}
\end{equation}
while the fish diagram in Fig.~\ref{box}\,(b), summing together with the other two channels, turns out to be finite,
\begin{equation}
{\cal A}_{\sigma}^{\rm fish}=-\frac{26m_D^4}{f_D^4\pi^2}\cdot\ln\left(\frac{4\pi^2\mu^2}{m_D^2}\right)\,.
\end{equation}

\subsection{The spin-2 glueballs}
In pure YM theory, the lightest state is known to be the spin-0 glueball, $0^{++}$, small fluctuations of the trace of the energy-momentum tensor of YM theory. Since the (anomalous) scale symmetry is spontaneously broken in the confined phase, the lightest $0^{++}$ state may be identified as the dilaton of pure YM theory~\cite{Hong:2017suj}. The lattice study further shows that the next-to-lightest state is the spin-2 state $2^{++}$ with a possible universal ratio of its mass with that of the lightest glueball~\cite{Bennett:2020hqd}. Therefore, as energy increases, the dilaton-dilaton scattering process will produce the spin-2 state, which may improve the ultraviolet divergence of the scattering amplitudes. 

Since the coupling of the spin-2 glueball with dilatons should preserve for consistency the diffeomorphism invariance, the spin-2 states couple to the energy-momentum tensor. The interaction Lagrangian density is given therefore in the leading order in the perturbation as
\begin{equation}
{\cal L}_{\rm int}=-\kappa \,h_{\mu\nu}T_D^{\mu\nu}\,,
\end{equation}
where $\kappa$ is the universal coupling of the spin-2 and $T_D^{\mu\nu}$ denotes the dilaton energy-momentum tensor.
The propagator of the massive spin-2 field with mass $m_G$ is given as 
\begin{equation}
\int_xe^{ip\cdot x}\left<0\right|T\left\{h_{\mu\nu}(x)h_{\alpha\beta}(0)\right\}\left|0\right>=\frac{iP_{\mu\nu\alpha\beta}}{p^2-m_G^2+im_G\Gamma_G+i\epsilon}\,,
\end{equation}
where $2P_{\mu\nu\alpha\beta}={\tilde\eta}_{\mu\alpha}{\tilde\eta}_{\nu\beta}+{\tilde\eta}_{\mu\beta}{\tilde\eta}_{\nu\alpha}-\frac{2}{3}{\tilde\eta}_{\mu\nu}{\tilde\eta}_{\alpha\beta}$ with ${\tilde\eta}_{\mu\nu}=\eta_{\mu\nu}-p_{\mu}p_{\nu}/m_G^2$ and $\eta_{\mu\nu}$ is the Minkowski metric. The decay width of the massive spin-2 state is given approximately in the leading order in $\kappa$ as~\cite{Agashe:2020wph} 
\begin{equation}
\Gamma_G\sim \kappa^2\frac{m_G^3}{960\pi}\,.
\end{equation}
The decay width is quite narrow and negligible even for a rather strong coupling $\kappa\, m_G\sim 1$.

\begin{figure}[t]
\centering
\begin{minipage}[c]{\linewidth}
\centering
  \includegraphics[scale=0.33]{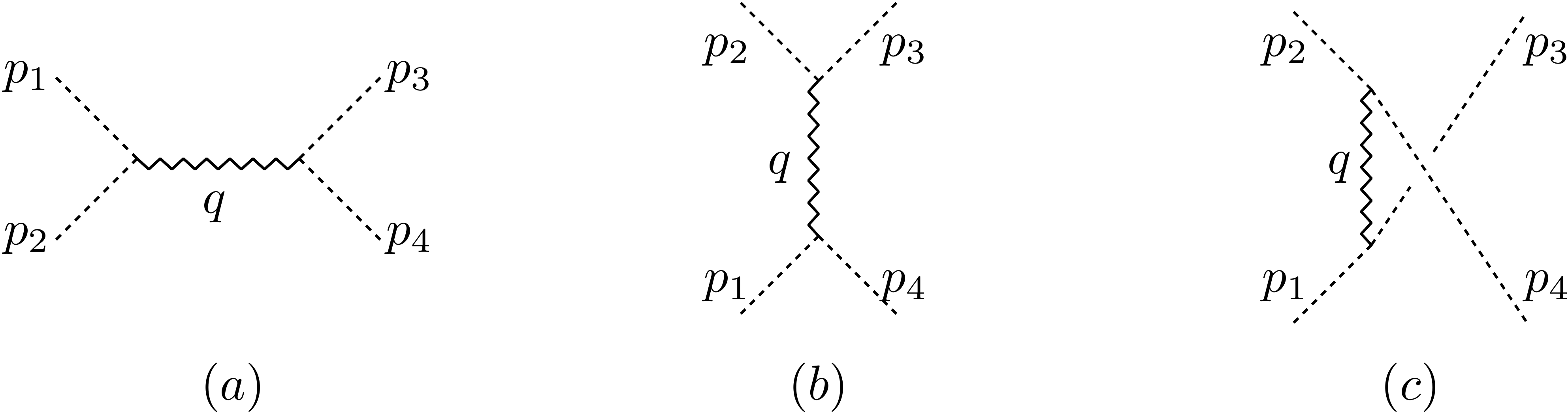}
\end{minipage}
\caption{Tree level diagrams for $2\to2$ scattering, mediated by the spin-2 glueball, denoted as spring-like lines: (a) $s$-channel, (b) $t$-channel, (c) $u$-channel. }
\label{graviton}
\end{figure}

Having the massive spin-2 glueball in the intermediate states, the dilaton scattering amplitude will be modified. The tree-level amplitude will have extra contributions mediated by the spin-2 glueball (Fig.~\ref{graviton}), which we find, neglecting the decay width,
\begin{equation}
{\cal A}_{\sigma}^{G}=-\frac{\kappa^2}{2}\left[\frac{-\frac23s^2+t^2+u^2}{s-m_G^2+i\epsilon}+\frac{s^2-\frac23t^2+u^2}{t-m_G^2+i\epsilon}+\frac{s^2+t^2-\frac23u^2}{u-m_G^2+i\epsilon}\right]\,.
\end{equation}
Combining all the diagrams, the tree, the one-loop and the spin-2 mediated, the dilaton scattering amplitude becomes for $s^2\gg t^2,u^2$, taking $m_D\approx0$ and $\Gamma_G\approx0$,
\begin{equation}
{\cal A}_{\sigma}\approx\frac{s^2}{16\pi^2f_D^4}+\frac{\kappa^2}{3}\frac{s^2}{s-m_G^2}\,,
\end{equation}
where we have absorbed into $f_D$ the logarithmic correction in 
Eq.~(\ref{box}).
\begin{figure}[t]
\centering
\begin{minipage}[c]{\linewidth}
\centering
  \includegraphics[scale=0.36]{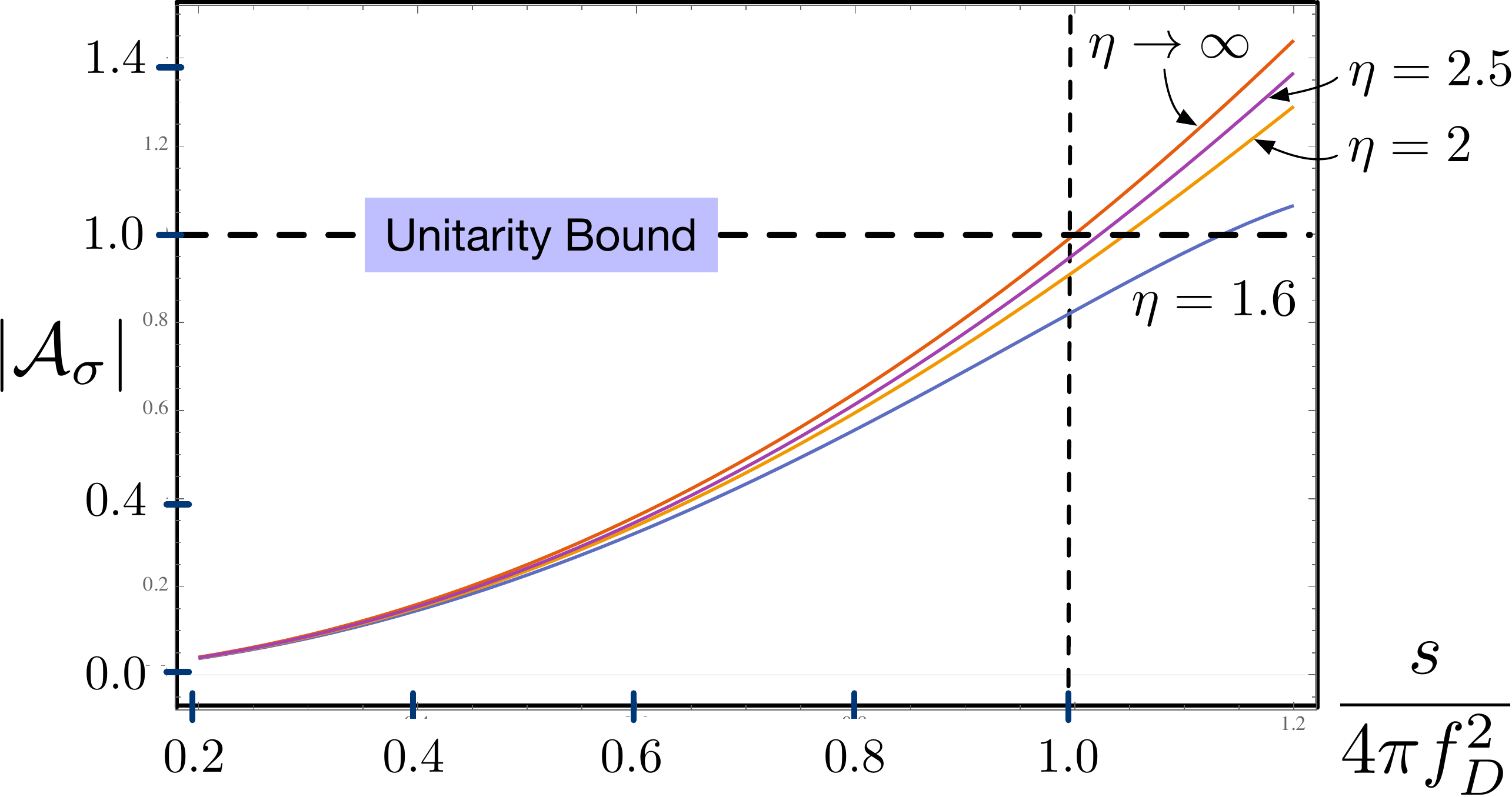}
\end{minipage}
\caption{The unitarity bound, $\left|{\cal A}_{\sigma}\right|\le1$,  for the $2\to2$ scattering amplitude of dilatons as a function of $s$ for $s^2\gg t^2,u^2$: Each plot corresponds to the inclusion of an intermediate spin-2 state of mass $m_G^2=4\pi\eta f_D^2$ for $\eta=1.6, 2, 2.5, \infty$ with the coupling $\kappa^2 =0.5/m_G^{2}$ as an example.  $\eta\to\infty$ corresponds to the decoupling limit.}
\label{unitarity}
\end{figure}
We plot the amplitude in Fig.~\ref{unitarity}. We see that the perturbative unitarity is violated at $s=4\pi f_D^2$ in the $2\to2$ dilaton scattering, if the spin-2 state is not included. Its inclusion however ameliorates the divergence~\footnote{In the case of $\pi\,\pi$ scattering. the inclusion of the spin-1 state, namely $\rho$ meson, improves the unitarity~\cite{Sannino:1995ik}
}.   Since the microscopic theory is unitary, there should appear the next-light states near the cutoff of the effective theory to restore the unitarity. As shown in Fig.~\ref{unitarity}, we find that  the unitarity is improved beyond the cutoff, when the intermediate spin-2 state of mass around $\sqrt{4\pi}f_D$ is included in the scattering process.

\subsection{Universal mass ratio}
We find that the spin-2 state of mass around the cutoff scale, $m_G\sim\sqrt{4\pi}f_D$, should appear to improve the unitarity violation in the dilaton effective theory. The unitarity argument then suggests the mass ratio to be 
\begin{equation}
	R\equiv \frac{m_{G}}{m_{D}}\sim \frac{f_D^2}{\left|{\cal E}_{\rm vac}\right|^{1/2}}\,.
\end{equation}
The vacuum energy density, ${\cal E}_{\rm vac}$, measures how big the breaking of the scale symmetry is, defining the infrared (IR) scale of the theory, $\Lambda_{\rm IR}$, while $f_D$ defines the scale, $\Lambda_{\rm SB}$, at which the scale symmetry is spontaneously broken\,\cite{Hashimoto:2010nw}.\footnote{The PCDC relation, $f_D^2m_{D}^2=-16\,{\cal E}_{\rm vac}$ associates the dilaton mass with the vacuum energy, the explicit scale-symmetry breaking term. When the explicit-breaking term becomes vanishingly small, the ground state is almost degenerate along the scale transformation and the dilaton becomes almost massless. Therefore, the dilaton decay constant should remain finite in that limit, which shows that they are two independent quantities.} In pure YM theory they are expected to be close to each other because it admits only a single scale, uniquely given by the renormalization group equation. Indeed the lattice study shows that $R\simeq 1.4$ for pure YM theory in $3+1$ dimensions while the models of its gravity dual give $\sqrt{2}\lesssim R\lesssim 1.74$~\cite{Bennett:2020hqd}.
In general, however, these two scales do not have to be of the same order. In fact, in theories of near conformal dynamics, they are predicted to be widely separated to follow the Miransky scaling or the Berezinskii-Kosterlitz-Thouless (BKT) scaling
\begin{equation}
	\Lambda_{\rm IR}=\Lambda_{\rm SB}\exp\left(-\frac{c}{\sqrt{\alpha_*/\alpha_c-1}}\right)\,,
\end{equation}
where $c$ is a ${\cal O}(1)$ constant and $\alpha_c$ is the parameter of the theory with $\alpha_*$ being the critical point of the phase transition. In the case of the Banks-Zaks theory~\cite{Banks:1981nn} $\alpha_*$ is the would-be IR fixed-point of the $\beta$ function, $\alpha_c$ is the critical coupling for the chiral symmetry breaking, the constant $c=\pi$, and the IR scale, $\left|{\cal E}_{\rm vac}\right|^{1/4}$, is given by the dynamically generated fermion mass due to the chiral symmetry breaking, $\Lambda_{\rm IR}\simeq m_{\rm dyn}$~\cite{Miransky:1996pd,Choi:2012kx,Hong:2013eta}
. By the suitable choice of the number of fermions and colors~\cite{Gies:2005as,Kaplan:2009kr}, or by turning on extra gauge interactions~\cite{Hong:2017smd}, one could make $\alpha_c$ very close to the critical point so that the separation of the scales can be as wide as possible. 

As anticipated in~\cite{Athenodorou:2016ndx}, we find that the unitarity argument shows that the mass ratio between the lowest spin-2 state and the dilaton measures the scale separation of the near conformal dynamics,
\begin{equation}
	R\equiv\frac{m_G}{m_{D}}\sim \exp\left(\frac{2c}{\sqrt{\alpha_*/\alpha_c-1}}\right)\,.
	\label{ratio}
	\end{equation}
If, therefore, one discovers both the new massive spin-2 state and the dilaton, one may be able to discern the correct conformal dynamics responsible for them. 

\subsection{Results and Discussion}
In this paper, we have calculated in perturbation the $2\to2$ dilaton scattering amplitudes from the dilaton effective theory to find that the one-loop amplitude violates the unitary bound at energy, $E\sim \sqrt{4\pi}f_D$. Such violation should be interpreted as a signal for an additional state, the next-lightest resonance, in the effective theory. We show that the spin-2 state, $2^{++}$, of mass around $\sqrt{4\pi}f_D$ improves the unitarity violation of the dilaton scattering amplitude near the cutoff of the effective theory.

The mass ratio between the dilaton and the spin-2 state, obtained from the unitarity argument, captures the degree of the conformality of the microscopic theory, Eq.\,(\ref{ratio}). The heavier the ground-state spin-2 is, compared to the dilaton, the more conformal the microscopic theory becomes. For instance, in pure YM theory, the lattice study shows the ratio is about 1.4~\cite{Bennett:2020hqd}, showing that the scale symmetry is badly broken in pure YM theory. This is expected, because in pure YM theory without fermions that screen the color charges, the $\beta$ function decreases from zero rapidly.  However, in theories like Banks-Zaks, where the $\beta$ function is almost zero for a wide range of scales near IR, the ratio $R$ could be quite large, if the chiral symmetry breaking occurs near the quasi IR fixed point, $\alpha_*-\alpha_c\ll\alpha_*$. 

\subsection{Acknowledgements}
We thank S.H. Im and J.-W. Lee for useful discussions. This work was supported by a 2-Year Research Grant of Pusan National University.
\subsubsection{}


\begin{thebibliography}{99}
\bibitem{Campostrini:1989uj}
M.~Campostrini, A.~Di Giacomo and Y.~Gunduc,
Phys. Lett. B \textbf{225}, 393-397 (1989)
doi:10.1016/0370-2693(89)90588-1
\bibitem{Ellis:1984jv}
J.~R.~Ellis and J.~Lanik,
Phys. Lett. B \textbf{150}, 289-294 (1985)
doi:10.1016/0370-2693(85)91013-5
\bibitem{Yamawaki:1985zg}
K.~Yamawaki, M.~Bando and K.~i.~Matumoto,
Phys. Rev. Lett. \textbf{56}, 1335 (1986)
doi:10.1103/PhysRevLett.56.1335
\bibitem{Bando:1986bg}
M.~Bando, K.~i.~Matumoto and K.~Yamawaki,
Phys. Lett. B \textbf{178}, 308-312 (1986)
doi:10.1016/0370-2693(86)91516-9

\bibitem{Hong:2017smd}
D.~K.~Hong,
JHEP \textbf{02}, 102 (2018)
doi:10.1007/JHEP02(2018)102
[arXiv:1703.05081 [hep-ph]].

\bibitem{Davoudiasl:2009cd}
H.~Davoudiasl, S.~Gopalakrishna, E.~Ponton and J.~Santiago,
New J. Phys. \textbf{12}, 075011 (2010)
doi:10.1088/1367-2630/12/7/075011
[arXiv:0908.1968 [hep-ph]].

\bibitem{Agashe:2020wph}
K.~Agashe, M.~Ekhterachian, D.~Kim and D.~Sathyan,
JHEP \textbf{11}, 109 (2020)
doi:10.1007/JHEP11(2020)109
[arXiv:2008.06480 [hep-ph]].

\bibitem{Hong:2017suj}
D.~K.~Hong, J.~W.~Lee, B.~Lucini, M.~Piai and D.~Vadacchino,
Phys. Lett. B \textbf{775}, 89-93 (2017)
doi:10.1016/j.physletb.2017.10.050
[arXiv:1705.00286 [hep-th]].

\bibitem{Bennett:2020hqd}
E.~Bennett, J.~Holligan, D.~K.~Hong, J.~W.~Lee, C.~J.~D.~Lin, B.~Lucini, M.~Piai and D.~Vadacchino,
Phys. Rev. D \textbf{102}, no.1, 011501 (2020)
doi:10.1103/PhysRevD.102.011501
[arXiv:2004.11063 [hep-lat]].

\bibitem{Migdal:1982jp}
A.~A.~Migdal and M.~A.~Shifman,
Phys. Lett. B \textbf{114}, 445-449 (1982)
doi:10.1016/0370-2693(82)90089-2

\bibitem{Choi:2012kx}
K.~Y.~Choi, D.~K.~Hong and S.~Matsuzaki,
JHEP \textbf{12}, 059 (2012)
doi:10.1007/JHEP12(2012)059
[arXiv:1201.4988 [hep-ph]].

\bibitem{Coleman:1969sm}
S.~R.~Coleman, J.~Wess and B.~Zumino,
Phys. Rev. \textbf{177}, 2239-2247 (1969)
doi:10.1103/PhysRev.177.2239

\bibitem{Callan:1969sn}
C.~G.~Callan, Jr., S.~R.~Coleman, J.~Wess and B.~Zumino,
Phys. Rev. \textbf{177}, 2247-2250 (1969)
doi:10.1103/PhysRev.177.2247


\bibitem{Komargodski:2011vj}
Z.~Komargodski and A.~Schwimmer,
JHEP \textbf{12}, 099 (2011)
doi:10.1007/JHEP12(2011)099
[arXiv:1107.3987 [hep-th]].

\bibitem{Sannino:1995ik}
F.~Sannino and J.~Schechter,
Phys. Rev. D \textbf{52}, 96-107 (1995)
doi:10.1103/PhysRevD.52.96
[arXiv:hep-ph/9501417 [hep-ph]].

\bibitem{Banks:1981nn}
T.~Banks and A.~Zaks,
Nucl. Phys. B \textbf{196}, 189-204 (1982)
doi:10.1016/0550-3213(82)90035-9

\bibitem{Miransky:1996pd}
V.~A.~Miransky and K.~Yamawaki,
Phys. Rev. D \textbf{55}, 5051-5066 (1997)
[erratum: Phys. Rev. D \textbf{56}, 3768 (1997)]
doi:10.1103/PhysRevD.56.3768
[arXiv:hep-th/9611142 [hep-th]].

\bibitem{Hong:2013eta}
D.~K.~Hong,
doi:10.1142/9789814566254\_0020
[arXiv:1304.7832 [hep-ph]].

\bibitem{Hashimoto:2010nw}
M.~Hashimoto and K.~Yamawaki,
Phys. Rev. D \textbf{83}, 015008 (2011)
doi:10.1103/PhysRevD.83.015008
[arXiv:1009.5482 [hep-ph]].

\bibitem{Gies:2005as}
H.~Gies and J.~Jaeckel,
Eur. Phys. J. C \textbf{46}, 433-438 (2006)
doi:10.1140/epjc/s2006-02475-0
[arXiv:hep-ph/0507171 [hep-ph]].

\bibitem{Kaplan:2009kr}
D.~B.~Kaplan, J.~W.~Lee, D.~T.~Son and M.~A.~Stephanov,
Phys. Rev. D \textbf{80}, 125005 (2009)
doi:10.1103/PhysRevD.80.125005
[arXiv:0905.4752 [hep-th]].

\bibitem{Athenodorou:2016ndx}
A.~Athenodorou, E.~Bennett, G.~Bergner, D.~Elander, C.~J.~D.~Lin, B.~Lucini and M.~Piai,
JHEP \textbf{06}, 114 (2016)
doi:10.1007/JHEP06(2016)114
[arXiv:1605.04258 [hep-th]].

\end{thebibliography}
\end{document}